\documentclass[twocolumn]{elsarticle}
\usepackage{graphicx}
\usepackage{multirow}
\usepackage{subcaption}
\usepackage[hyphens]{url}
\usepackage{amsmath}
\usepackage{mathtools}
\usepackage{float}

\begin{document}
\twocolumn[{

\begin{frontmatter}

\title{Autonomic Management of Power Consumption with IoT and Fog Computing}

\author{Hugo Vaz Sampaio}
\author{Fernando Koch\corref{corresponding}}
\cortext[corresponding]{Corresponding author}
\ead{fkoch@acm.org}
\author{Carlos Becker Westphall}
\author{Ricardo do Nascimento Boing}
\author{Rene Nolio Santa Cruz}

\address{UFSC–CTC–INE–LRG PO. Box 476, 88040-970, Florianópolis, SC, Brazil}

\begin{abstract}

We introduce a system for Autonomic Management of Power Consumption in setups that involve Internet of Things (IoT) and Fog Computing. The \textit{Central IoT} (CIoT) is a Fog Computing based solution to provide advanced orchestration mechanisms to manage dynamic duty cycles for extra energy savings. The solution works by adjusting Home (H) and Away (A) cycles based on contextual information, like environmental conditions, user behavior, behavior variation, regulations on energy and network resources utilization, among others. Performance analysis through a proof of concept present average energy savings of $58.4\%$, reaching up to $61.51\%$ when augmenting with a scheduling system and variable long sleep cycles (LS). However, there is no linear relation increasing LS time and more savings. The significance of this research is to promote autonomic management as a solution to  develop more energy efficient buildings and smarter cities, towards sustainable goals.

\end{abstract}


\end{frontmatter}
}]

\section{Introduction}
\label{sec:intro}

Home Energy Management Systems (HEMS) are popularly used to monitor energy consumption,identify deviations, and promote optimization of energy utilization. These solutions are usually built by integrating data collected from sensors, user behaviour, environmental information, Micro-Grid networks, Heating Ventilation Air Conditioning (HVAC), renewable energy sources, and other data sources and historical information to to subsidize strategies for energy preservation. Examples include in popular products like Google Nest Learning Thermostat, EcoBee Smart Thermostat, and others.

We devised a solution to optimize energy consumption by promoting adjustments to regulate energy utilization on a regular basis. The \textit{Central IoT} (CIoT) is a Fog Computing based solution to provide advanced orchestration mechanisms to manage dynamic duty cycles for extra energy savings. The solution works based on IoT devices able to identify periods of time when residents are at home (state H) or away (state A). The method works by adjusting H and A cycles based on contextual information, like environmental conditions, user behavior, behavior variation, regulations on energy and network resources utilization, among others. We foresee application scenarios around e.g. smart homes, smart cities, smart agriculture, intelligent fire alarms, intelligent traffic lights, and others. 

We are taking into consideration the challenges and demands for meeting real-time system requirements, guaranteeing large scale system network stability, and a lightweight management protocol. These solutions will be useful in scenarios such as traffic management in smart cities~\cite{Mostafa2017}, smart greenhouses in agriculture~\cite{Sampaio2017a, Sampaio2017b}, intelligent fire alarms in smart homes~\cite{sampaio2019}, for instance. The key contributions of this paper include: 

\begin{enumerate}
  
   \item A proposal for an advanced orchestration mechanism for IoT devices in Fog Computing environment aiming to optimize power consumption.
         
   \item A proof-of-concept implementation of large scale IoT environments and analysis of the results.
   
   \item An analysis of the state-of-the-art weighting the contribution in perspective.
\end{enumerate}

The paper is structured as follows. Section \ref{sec:proposal} details the proposal. Section \ref{sec:experiments} present the experiments implemented to validate the proposal. Then, Section \ref{sec:results} analyse the results. Section \ref{sec:background} describe some background and related work and the paper concludes with a summary of our findings and future work.


\section{Proposal}
\label{sec:proposal}

\begin{figure}[!ht]
\centering
\includegraphics[width=.49\textwidth]{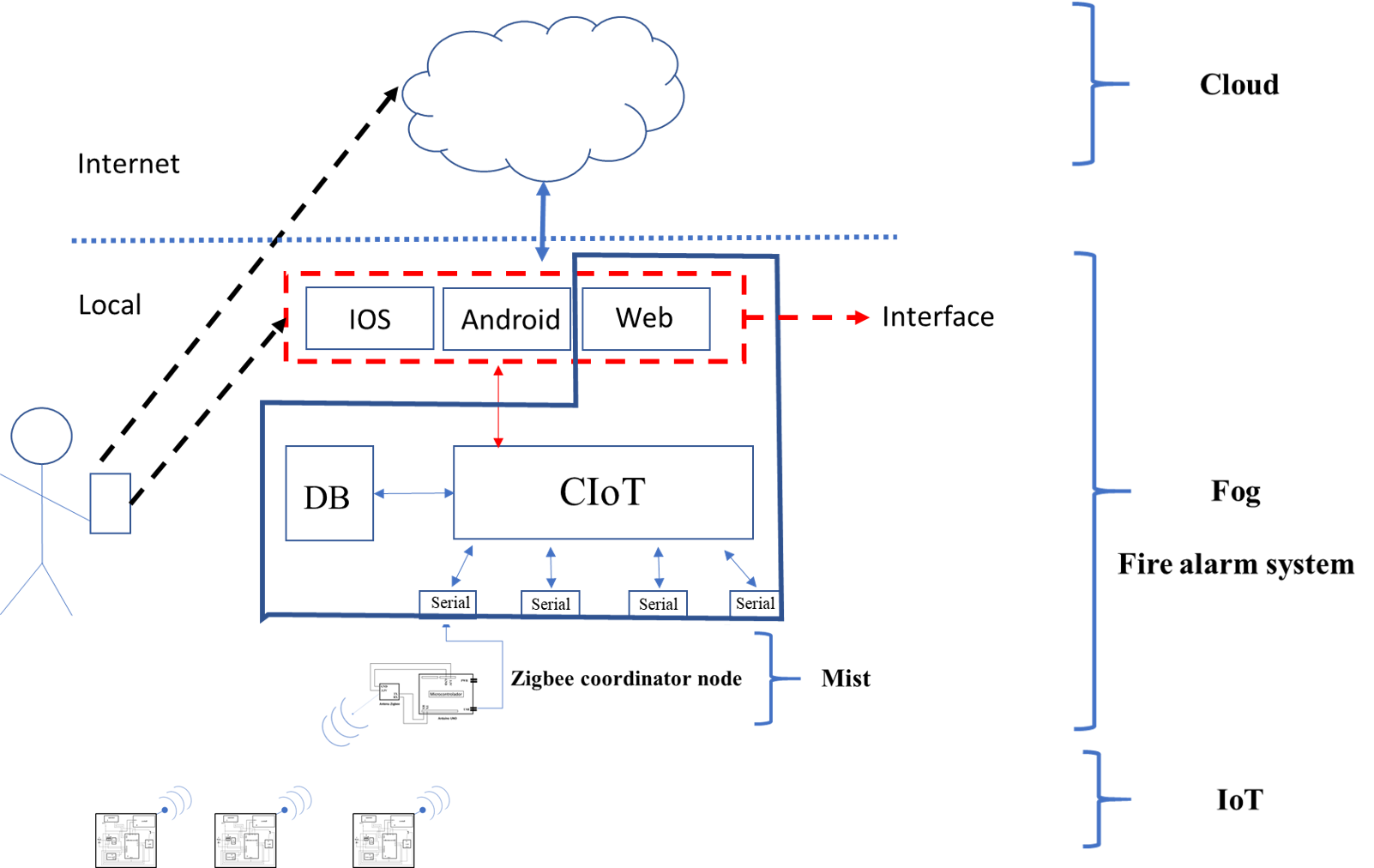}
\caption{Proposed Solution Architecture of Central Internet of Things (CIoT)}
\label{fig:Fog_diagram}
\end{figure}

Fig. \ref{fig:Fog_diagram} depicts the proposed solution architecture encompassing 3 layers: IoT devices, Fog layer, and Cloud layer. The central module is responsible for managing a local application with a web and database servers, as well as controlling communications with IoT devices. We consider two possible distribution scenarios: 

\begin{enumerate}

\item \emph{Fog Configuration}, where the radio coordinator module is embedded with a Fog device and communicate directly with IoT devices.

\item \emph{Mist Configurations}, where the  radio coordinator node is connected as a Mist layer to a Fog serial USB port, such as using a ZigBee protocol.

\end{enumerate}

Let us think of a illustrative use scenario as fire alarm system that automatically updates its own sensors' based on the current state -- i.e. Regular Operation, Emergency Operation, Away Mode -- and provided user schedule. 

In a Fog Configuration, the fire alarm system application can be accessed either locally or through the Internet, along with estimated energy utilization by the devices. The average packet size contain 25 bytes. 

The Fog server controls the fire alarm IoT devices by altering the sleep time values, turning the device's alarm and sensors on and off. The control messages provide four functions and their respective options, with the first field representing which type of control to modify, and the second field defining the possible values of the control type. 

In a Mist configuration, the ZigBee coordinator receives data from IoT nodes, compresses the data packets, and forwards them to the Fog element. Here, the average packet size of 10 bytes, including:  Fire Alarm Identifier byte (0x11) hex value, 2 bytes of  source address, and 7 bytes of sensors values.Reducing the source address from the ZigBee packet causes the Mist node to maintain an address translation table for communication between the Fog and IoT devices. 

With the Mist layer device, the Fog system no longer communicate directly with the IoT device. The control frames are sent from the Fog node to the Mist node, so, CIoT generates a packet with an initial identifying byte (0x11) hex value, followed by two bytes of IoT address destination. Next, the control frame is added to the control packet and forwarded to the Fog device's serial port. The Mist device then builds and forwards the ZigBee packet to the IoT device. 

\begin{figure}[ht!]
\centering
  \includegraphics[width=.9\linewidth]{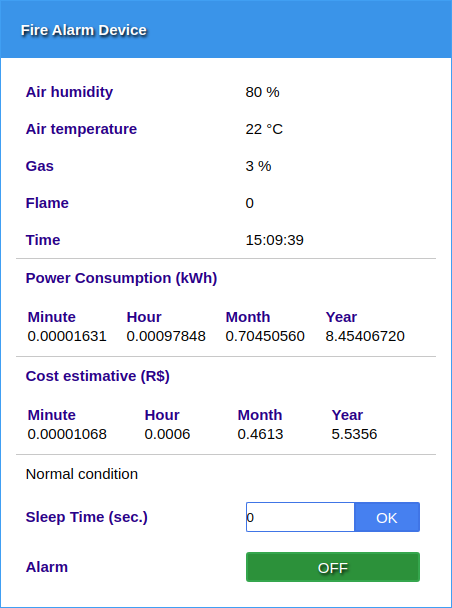}
\caption{IoT Fire Alarm Web Management Interface.} \label{fig:systeminterface}
\end{figure}

The IoT devices are managed through a web application, exemplified in Fig. \ref{fig:systeminterface}. Type \textbf{"a"} messages are generated when the alarm button is pressed, with two options: \textbf{"L"} to turn on the device alarm and \textbf{"D"} to turn off. Type \textbf{"b"} messages alter the sleep time with an input value between 0 and 3, altering the device's duty cycle. Power consumption and cost estimates are updated automatically.

\section{Experiments}
\label{sec:experiments}

\begin{figure}[ht!]
    \centering
    \includegraphics[width=.48\textwidth] {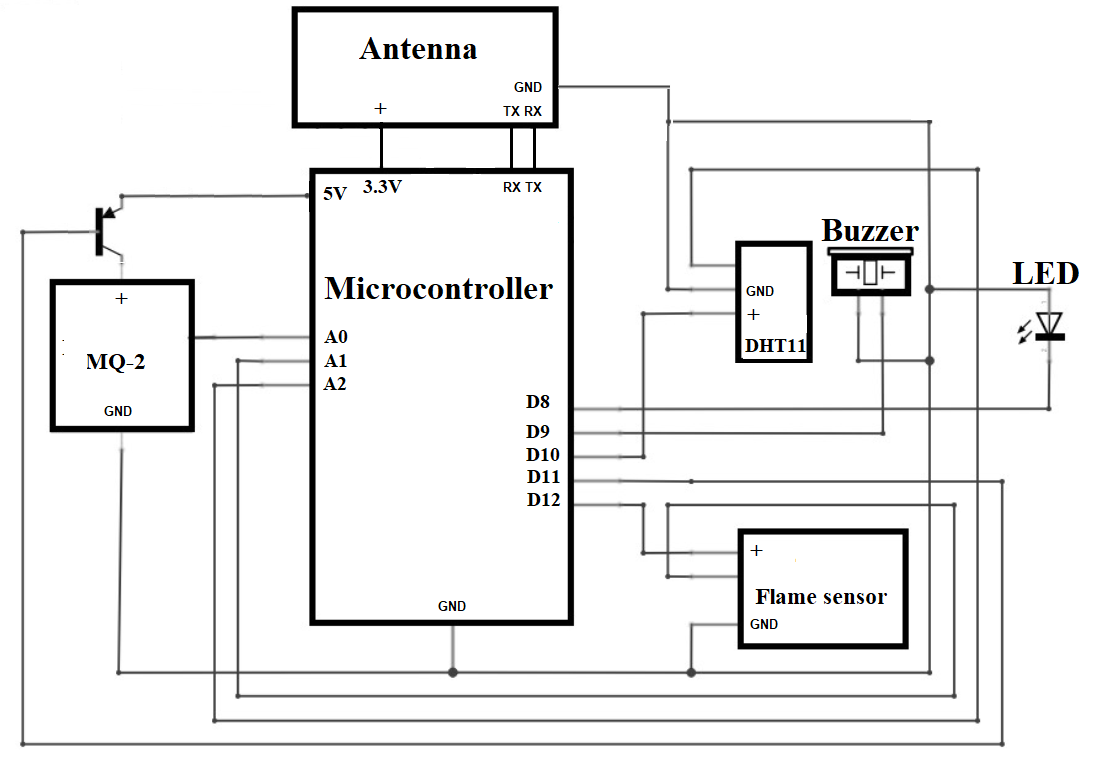}
    \caption{IoT Fire Alarm Device's Physical Connections.}
    \label{fig:IoTalarm}
\end{figure}

We propose two experiments to estimate the energy savings for these configurations: (i) we estimate the energy consumption of the networked devices by varying the sleep time in a household environment, and (ii) we expand the scenario to a condominium context with multiple households. In both experiments, we use a powered fire alarm IoT node device, depicted in Fig.\ref{fig:IoTalarm}. 

The IoT device is composed of a 8-bit Microcontroller (MC), an antenna module (ZigBee), air (DHT11), Gas (MQ-2) and flames sensors, sound (buzzer) and light (LED) actuators. To determine the IoT's fire alarm duty cycle, we limited the Regular operation active cycle, to the sensors longest response time (RT), being the Air sensor module's RT of 2s, while gas and flames sensors' RT is 1s.

The radio module connects to the microcontroller using only 2 wires, for transmitting (TX) and receiving (RX) data, with a serial UART half-duplex communication interface. Radio modules with different protocols, such as WiFi, Bluetooth, and LoRa, could also be used with the UART interface. In this configuration, the radio module and microcontroller communication, is limited at default baud rate of $115200bps$. 

\begin{table}[!ht]
\centering
\caption{Modules' Current (mA) and Response Time (RT)(s) Performance Metrics in Regular Operation.}
\label{tab:IoT_modules}
\resizebox{.48\textwidth}{!}{
\begin{tabular}{c|c|c|c|c}

i & Module &  Ch (mA) & Cs (mA) &RT (s).\\

\hline
1 & DHT11 sensor &  0.3 & $8.33*10^{-5}$ & 2\\
2 & MQ-2 sensor &  160 & $4.44*10^{-2}$ & 1\\
3 & Flames sensor & 0.4 & $1.11*10^{-4}$ & 1\\
4 & Arduino microprocessor (Active mode) & 0.3 & $8.33*10^{-5}$ & 2\\
5 & XBee ZigBee antenna (Send mode) &  33 & $9.17*10^{-3}$ & 0,0008\\
6 & XBee ZigBee antenna  (Receive mode) & 28 & $7.78*10^{-3}$ & 1,9992\\
7 & Arduino microprocessor (sleep-mode) & $10^{-5}$ & $2.78*10^{-8}$ & T\\
8 & XBee ZigBee antenna (sleep-mode) & $10^{-4}$ & $2.78*10^{-8}$ & T\\
\end{tabular}
}
\end{table}

\begin{table}[!ht]
\centering
\caption{Modules' Current (mA) and Response Time (RT)(s) Performance Metrics in Emergency Operation.}
\label{tab:IoT_modules_emergency}
\resizebox{.48\textwidth}{!}{
\begin{tabular}{c|c|c|c|c}

i & Module &  Ch (mA) & Cs (mA) & RT (s).\\
\hline
1 & DHT11 air sensor &  0.3 & $8.33*10^{-5}$ & 1\\
2 & MQ-2 gas sensor &  160 & $4.44*10^{-2}$ & 1\\
3 & Flames sensor & 0.4 &$1.11*10^{-4}$ & 1\\
4 & Arduino microprocessor (Active mode) & 0.3 & $8.33*10^{-5}$ & 1\\
5 & XBee ZigBee antenna (Send mode) &  33 & $9.17*10^{-3}$ & 0,0008\\
6 & XBee ZigBee antenna  (Receive mode) & 28 & $7.78*10^{-3}$ & 0,9992\\
7 & Buzzer &  25 & $6.94*10^{-3}$  & 1\\
8 & LED &  20 & $5.56*10^{-3}$ & 1\\

\end{tabular}
}
\end{table}

Table \ref{tab:IoT_modules} presents the performance metrics for Regular Operation as provided by the manufacturer, whereas table \ref{tab:IoT_modules_emergency} depicts the metrics for Emergency Mode when the device is in constant watch sending one packet per second to the Fog Computing infrastructure. In both cases, the metrics are provided in milliAmperes (mA).

During Regular active cycle, we consider that the radio module will stay on Receive mode, thus available to receive control or emergency messages from CIoT. At the end of the active cycle, the sensors values are read by MC, then the radio forwards the packet to CIoT, all sensors are turned off, and the antenna and MC are put into sleep mode. 

However, we notice that in practice there are regulatory limits associated with duty and sleep cycles. For instance, the normative instruction IN012/DAT/CBMSC of the State Fire Brigade in the State of Santa Catarina, Brazil, defines that electronic devices used for fire alarms must have a maximum response time of 5 seconds. If the duration of the active cycle is 2 seconds, then the maximum sleep mode is 3 seconds.

The duty cycle energy consumption $EC$ for each networked device can be expressed as the Eq.\ref{energy_sum}, where $Cs_i$ is the power utilization and $RT_i$ is the response time for each of the $n$ device's modules:

\begin{equation}
   EC = \sum_{i=1}^{n} (Cs_i \times RT_i)
   \label{energy_sum}
\end{equation} 

The energy consumption  or a fire alarm node operating in Regular Operation with duty cycles of 2 seconds can be calculated as below, based on the metrics provided in table \ref{tab:IoT_modules}:

\begin{equation}
\begin{multlined}
   EC = (0.3 \times 8.33*10^{-5} \times 2) + \\
   (160 \times 4.44*10^{-2} \times 1) + \\
   (0.4 \times 1.11*10^{-4} \times 1) + \\
   (0.3 \times 8.33*10^{-5} \times 2) + \\
   (33 \times 9.17*10^{-3} \times 0,0008) + \\
  (28 \times 7.78*10^{-3} \times 1,9992) = 0.0604mA
  \end{multlined}
\end{equation}

Hence, in a configuration where fire alarms operate with 9V, the Total Power Consumption $TWC$ for each cycle can be calculated as:

\begin{equation}
  TWC = 9V \times 0.0604 \textrm{mA} = 0.5436 \textrm{mW/cycle}
  \label{Power_cycle_consumption}
\end{equation}

Correspondingly, we calculate the consumption per day, month and year for different variations of sleep time T in table \ref{tab:powerconsumption}. 

\begin{table}[!ht]
\centering
\caption{One IoT Device Power Consumption, in Regular Mode, in kWh per Minute, Hour, Day, Month and Year, With Variations of Sleep Time T}
\label{tab:powerconsumption}
\resizebox{.5\textwidth}{!}{
\begin{tabular}{c|c|c|c|c|c|c|c|c}
T &
  Current (mA) &
  Cycle (mW) &
  \textbf{minute} &
  \textbf{hour} &
  \textbf{day} &
  \textbf{month} &
  \textbf{year} &
  S(\%) \\ \hline
0 & 0.0604    & 0.543600 & $1.63*10^{-5}$ & $9.78*10^{-4}$ & $2.35*10^{-2}$ & 0.705 & 8.45 & -       \\
1 & 0.0604003 & 0.543602 & $1.09*10^{-5}$ & $6.52*10^{-4}$ & $1.57*10^{-2}$ & 0.470 & 5.64 & 33.33   \\
2 & 0.0604006 & 0.543605 & $8.15*10^{-6}$ & $4.89*10^{-4}$ & $1.17*10^{-2}$ & 0.352 & 4.23 & 49.99   \\
3 & 0.0604009 & 0.543608 & $6.52*10^{-6}$ & $3.91*10^{-4}$ & $9.4*10^{-3}$ & 0.282 & 3.38 & 59.99   \\
\end{tabular}
}

\end{table}

Therefore, we conclude that each fire alarm device will consume about 8.4 kWh/year if operating in Regular Mode without sleep time (T=0). However, by setting up a regular sleep time of 3s (T=3), the device will consume 3.38 kWh/year, representing an optimization in power consumption by 59.99\%  whilst still complying with regulations. 

Moreover, we analyze the consumption and savings for different variations of $EC_{emergency}$ mode scenarios. We estimate the device's emergency mode energy consumption with Eq. \ref{energy_sum} and table \ref{tab:IoT_modules_emergency}, with a duty cycle of 1 second, resulting in $EC_{emergency}$ = 0.065mA. Therefore, while operating in Regular mode with T=3, we estimate savings when device operates in Emergency mode for 1\%, 2\%, 5\% and 10\% of the time. With devices operating in Emergency mode for 1\% of the time, maximum energy savings will result in 58.25\%, or 1.74\% lower than sole use of Regular mode with T=3.

\subsection{Extended Scenario}

Let us consider the context of an apartment building complex, where fire alarm devices connect to a \textit{centralized monitoring node} in a Fog Computing configuration. In this context, some systems will be shared by multiple residences, such as emergency alarm systems, access control, security, and consumption management of water and electricity. 

Moreover, there will be some level of collaboration: e.g. if a fire is detected in an apartment, the monitoring node will broadcast to all units to activate their audible and visual alarms, thus informing all neighboring residents about the state of emergency. To enable the collaboration features, in a configuration where the duty cycle is 2s, radio will remain listening for 1.9992s to received broadcast communication, being the remainder time the window for transmission.

\begin{figure}[ht!]
\centering
\includegraphics[width=0.49\textwidth]{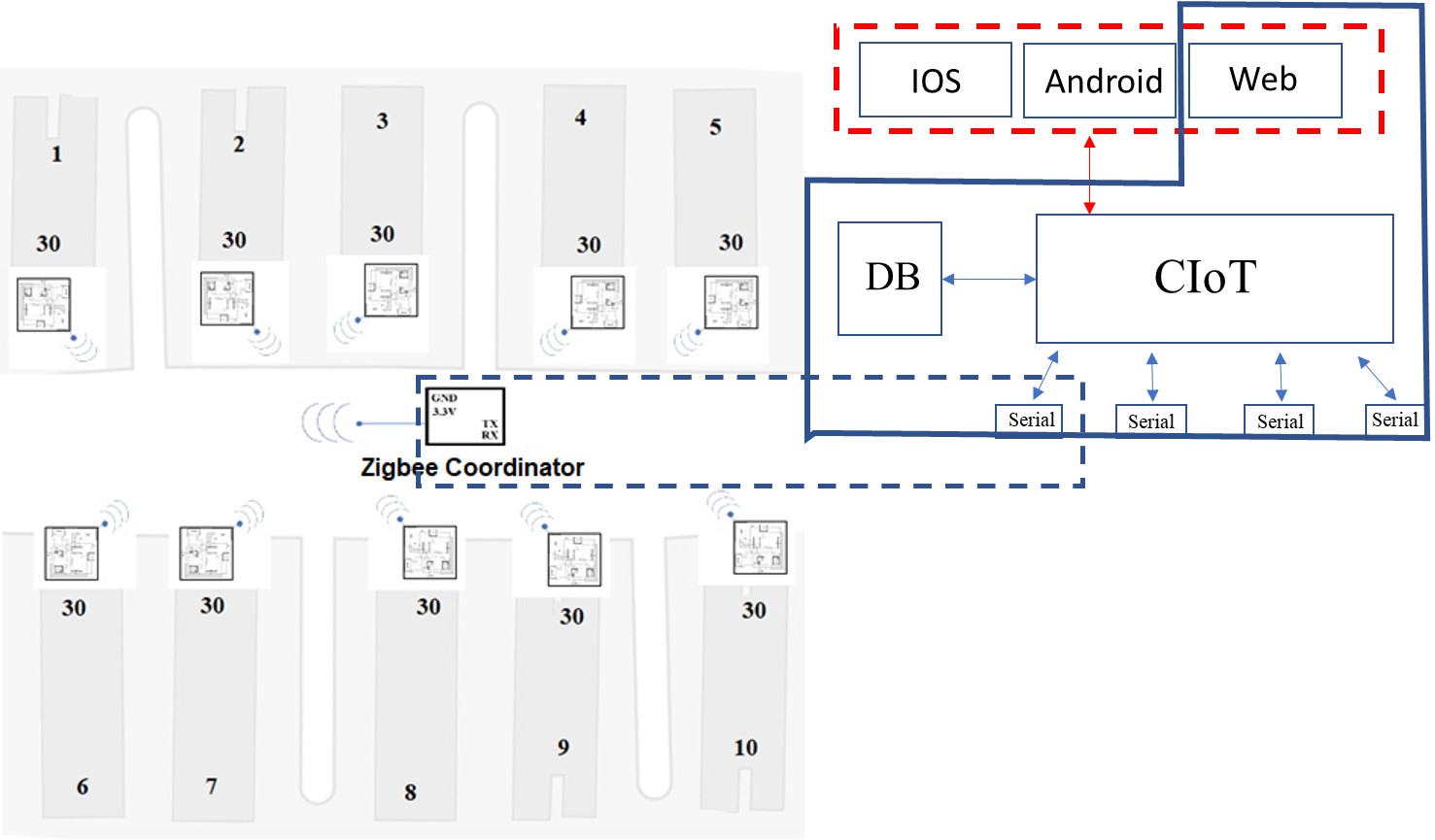}
\caption{Proposed Smart Condominium Fire Alarm System Example}
\label{fig:condominium_Fog}
\end{figure}

\begin{table}[!ht]
\centering
\caption{300 IoT Devices Estimated Power Consumption (kWh) per minute, hour, day, and month in a Condominium with 300 Apartments}
\label{tab:consumo300}
\resizebox{0.48\textwidth}{!}{%
\begin{tabular}{c|cccccc}
T & cycle (mW) & minute & hour & day & month & year \\ \hline
0 & 163.0800 & $4.89*10^{-3}$ & 0.29 & 7.05 & 211.35 & 2536.22 \\
1 & 163.0808 & $3.26*10^{-3}$ & 0.19 & 4.70 & 140.90 & 1690.82 \\
2 & 163.0817 & $2.45*10^{-3}$& 0.15 & 3.52 & 105.68 & 1268.12 \\
3 & 163.0825 & $1.96*10^{-3}$& 0.12 & 2.82 & 84.54 & 1014.50 \\
\end{tabular}%
}
\end{table}

Let us consider a building complex with 10 buildings and 30 apartments per building, totaling 300 apartments, with 1 fire alarm device per apartment, thus 300 IoT devices, as depicted in Fig. \ref{fig:condominium_Fog}. The energy consumption per minute, hour, day, month, and year, for different variations of sleep time $T$, are depicted in table \ref{tab:consumo300}. For T equal to 0, the setup will consume approximately 2.5 MW/year; however, with sleep time of 3s (T=3), the power consumption is optimized by 1.5 MW/year, representing a significant saving in power resources. Similarly, we analyze the consumption and savings for different variations of Emergency Mode scenarios in table \ref{tab:condominiumemergencyconsumption}. With condominium devices operating in Emergency mode for 1\% of the time, we estimate an energy consumption difference of 44.45kW per year. 

\begin{table}[!ht]
\centering

\caption{300 IoT Devices Power Consumption (kWh) in Emergency Mode.}
\label{tab:condominiumemergencyconsumption}
\resizebox{.5\textwidth}{!}{
\begin{tabular}{c|c|c|c|c|c|c|c}
Emergency & Cycle (mW) & minute & hour & day & month & year & Savings (\%) \\ \hline
1\%       & 163.21     & 2.04E-03   & 0.12     & 2.94    & 88.25     & 1058.95  & 58.25        \\
2\%       & 163.33     & 2.13E-03   & 0.13     & 3.06    & 91.95     & 1103.39  & 56.49        \\
5\%       & 163.70     & 2.39E-03   & 0.14     & 3.44    & 103.06    & 1236.72  & 51.24        \\
10\%      & 164.32     & 2.81E-03   & 0.17     & 4.05    & 121.58    & 1458.93  & 42.48       
\end{tabular}
}
\end{table}

\section{Results}
\label{sec:results}

In this section, we analyze the fire alarm system's IoT network performance, considering a hierarchical tree configuration, using Queuing Theory. In this setting, each IoT device uses its ZigBee antenna to forward information to a centralized ZigBee Coordinator node in the Fog, creating a processing bottleneck at the antenna-to-microcontroller serial connection in the system. In Sampaio and Motoyama \cite{Sampaio2017a, Sampaio2017b}, an analysis of a large-scale sensor network system is applied to agricultural greenhouses. The results indicated that system scalability is possible, with support for thousands of sensor nodes. However, the parameters considered determined a 5-minute interval between packet sendings, as well as an average queue system time, in the coordinator, of 2 seconds.

Let us consider that each IoT node will send one packet per duty cycle, of two seconds when T equals 0, and five seconds when T equals 3. When scaling the network to 300 IoT nodes, the delay time related to the transmission between the antennas, should also consider packets collision and re-transmission delay due to the network density (D). In this paper, we focus our analysis on the processing the total IoT ZigBee packets input flux at coordinator $\lambda_C$ (lambda), being the packets that will effectively arrive at the coordinator's antenna queue, can be considered a Poisson distribution. Thus, we find the average packets per second, show in Eq. \ref{eq:150} : 

\begin{equation}
\label{eq:150}
    \lambda_C = 300 \textrm{pct/2s} = 150 \textrm{pct/s}. 
\end{equation}

In this case, the Fog ZigBee antenna communicates via UART (U) serial port, with a speed limit in bits per second (bps) of $U = 115200\textrm{bps}$. The considered ZigBee average packet size is $P = 25$ bytes or 200 bits. The service rate $\mu_C$ (mu), at output of coordinator, is the serial port connection speed U divided by average packet size P:   

\begin{equation}
   \mu_C = \frac{U}{P} = 576 \textrm{pct/s}
\end{equation}

The load $\rho$ (rho), in Eq. \ref{eq:rho}, is defined in general as the ratio between incoming packet rate $\lambda$, and output packet rate $\mu$.

\begin{equation}
\label{eq:rho}
    \rho_C = \frac{\lambda_C}{\mu_C}
\end{equation}

Each packet average time spent in the system is $E\{T\}$, considering the waiting time added with packet service time \cite{Sampaio2017a}. We now calculate the condominium $E\{T\}$ multiplying by $\lambda_C$ in Eq. \ref{eq:et}: 

\begin{equation}
\label{eq:et}
    E\{T\} = \frac{1}{\mu_C - \lambda_C} \times \lambda_C
\end{equation}

\begin{figure}[!ht]
\centering
\includegraphics[width=0.49\textwidth]{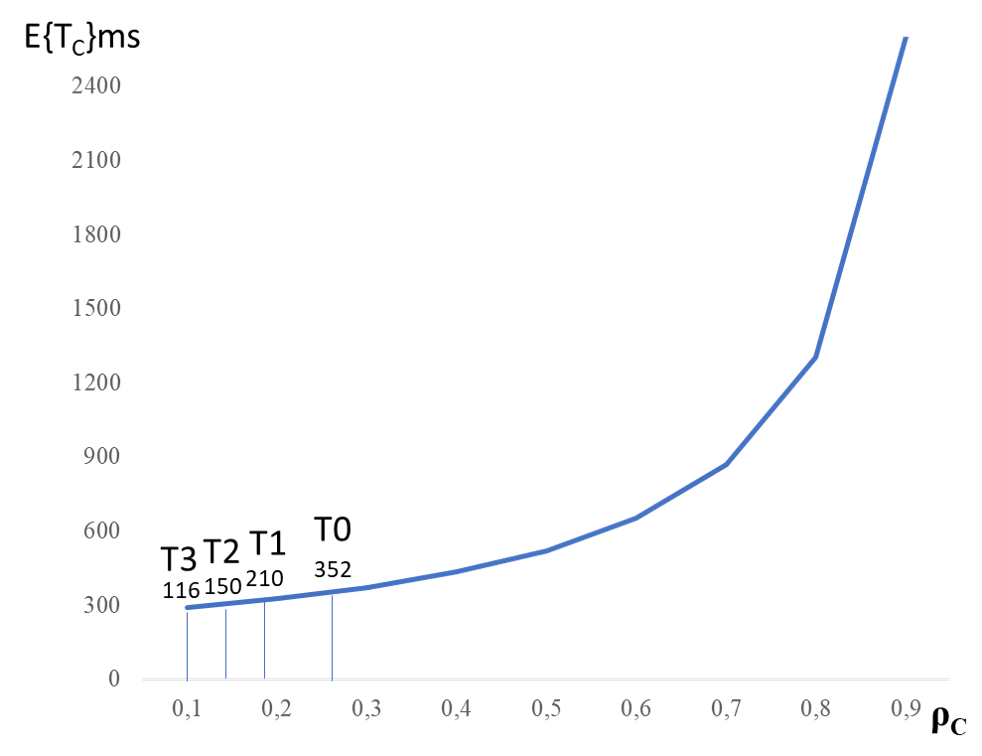}
\caption{Average waiting time at coordinator $E\{T_C\}$ and load $\rho_C$ variation curve}
\label{fig:rhoFog}
\end{figure}

Thus, as the load $\rho_C$ at coordinator C rises, the packet average time spent in the queue will also rise. Fig. \ref{fig:rhoFog} displays a curve graph of average waiting time at coordinator $E\{T_C\}$ with $\lambda_C$ variation. With $T = 0$, input flux is $\lambda_C$ = 150 pct/s, $\rho_C = 0.26$, resulting in 352ms average waiting time. Variation of T will alter $\lambda_C$, $E\{T_C\}$ and $\rho_C$, shown in table \ref{T_coordinator}.

By varying all IoT alarm devices' sleep time T from 0 to 3 seconds in table \ref{T_coordinator}, it will alter average packets input $\lambda_C$ from 150 to 60 packets per second, decreasing average waiting time $E\{T_C\}$ from 352 to 116 milliseconds, and decrease $\rho_C$ occupancy level from 0.260 to 0.104. Thus, as $\lambda_C$ decreases it results in higher savings.

\begin{table}[!ht]
\centering
\caption{Time T, Average Packets Input $\lambda_C$ and Estimated Savings (\%S) variation}
\resizebox{.33\textwidth}{!}{
\begin{tabular}{ccccc}
\textbf{T} & $\lambda_C$ & $E\{T_C\}$ms & $\rho_C$ & S (\%) \\ \hline
\multicolumn{1}{c|}{0} & 150               & 352              & 0.260  & -    \\
\multicolumn{1}{c|}{1} & 100               & 210              & 0.174  & 33.33    \\
\multicolumn{1}{c|}{2} & 75                & 150              & 0.130  & 49.99     \\
\multicolumn{1}{c|}{3} & 60                & 116              & 0.104 &  59.99     
\end{tabular}}
\label{T_coordinator}
\end{table}

We also consider the fire alarm IoT devices to be managed remotely, thus, CIoT is responsible for generating the devices' controls and emergency messages, considered as Feedback packets. In this case we use Jackson's theorem, from Queuing Theory, to calculate $E\{T_C\}$. This theorem establishes that in a network of queues with feedback, having negative exponential service, and an input flux from outside source, each queue can be treated as an M/M/m independent queue.

\begin{figure}[!ht]
\centering
\includegraphics[width=0.48\textwidth]{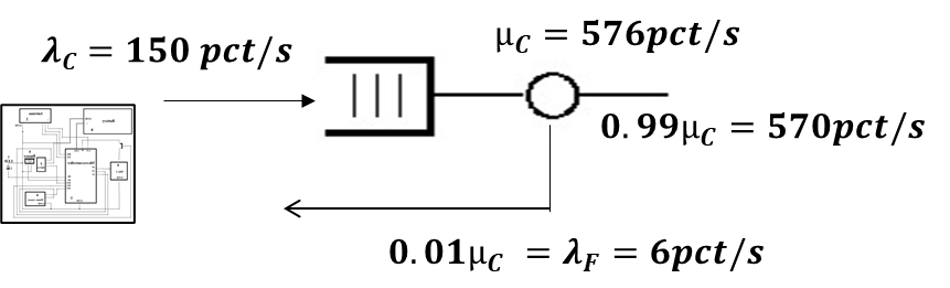}
\caption{Fog queue system with input $\lambda_C$, processing rate $0.99\mu_C$ and feedback rate $\lambda_F$.}
\label{fig:feedbackFog}
\end{figure}

Consequently, we estimate energy savings considering a reserve for feedback packets named $\lambda F_C$. The Fog system queue is shown in Fig. \ref{fig:feedbackFog}. For example, with $0.01\mu_C$ as feedback, it alters the receiving packets processing rate to 0.99$\mu_C$ = 570 pct/s. With 0.99$\mu_C$ and T=0, $E\{T_C\}$ will be 356ms, resulting in an estimated delay difference of 4ms, and with T=3 the difference is 1ms. 

The packet network average waiting time $E\{T_C\}$ must then be considered within the IoT duty cycle and its 3 seconds sleep limit. We then add the condominium $E\{T_C\}$ with T, resulting in the sleep Total Time (TT) to maximize condominium energy savings, shown in Eq. \ref{waiting_t}. 

\begin{equation}
    E\{T_C\} + T = TT \leq 3s
    \label{waiting_t}
\end{equation}

Table \ref{T_lambda_feedback} exemplifies TT values for $E\{T_C\}$ when also considering feedback. In case without feedback, $T = 2.88$ maximizes estimate condominium energy savings to 59\%, when compared to $T = 0$. With 1\% feedback there is 0.1\% savings difference, while with 10\% feedback, savings difference would be 0.2\% lower, or 58.8\%. 

\begin{table}[!ht]
\centering
\caption{IoT Network With Feedback Estimated Savings S(\%)}

\resizebox{.49\textwidth}{!}{
\begin{tabular}{cccccccc}
Feedback & \textbf{T} & $\lambda_C$ & $\lambda_{FC}$ & $E\{T_C\}$(s) & $\rho_C$ & TT(s) & S(\%)\\ 
\hline
No feedback & 2.88 & 61.5 & - & 0.1195 & 0.1067 & 2.999 & 59   \\ 
$0.01\mu_C$ & 2.87 & 61.6 & 6 & 0.1209 & 0.1079 & 2.991 & 58.9   \\ 
$0.05\mu_C$ & 2.87 & 61.6 & 29 & 0.1269 & 0.1126 & 2.997 & 58.9   \\ 
$0.10\mu_C$ & 2.86 & 61.7 & 58 & 0.1353 & 0.1192 & 2.995 & 58.8   \\ 

\end{tabular}}
\label{T_lambda_feedback}
\end{table}

\subsection{Network performance analysis with Mist}

In this Section, the system analysis is performed with addition of a Mist layer. In this new configuration, the Mist node device, composed of an Arduino and a ZigBee antenna, is the ZigBee coordinator node, with average waiting time $E\{T_C\}$. Packets are then reduced in the Mist node and forwarded to the Fog, this being the Mist to Fog queue, with average waiting time $E\{T_M\}$. In queue networks, Burke's theorem states that the exit of each queue, in tandem with Poisson entry and negative exponential service, is also Poisson, and can be considered as independent M/M/m queues.

With Mist and Fog queues in tandem, IoT packets leaving the Mist node will also be considered Poisson upon arriving at Fog. By treating both queues as independent, the total average time $E\{T_{CM}\}$ will be the sum of the Mist queue $E\{T_C\}$ and the Fog queue $E\{T_M\}$, being simplified to one queue.

IoT packets are received by the Mist node, with a maximum average rate of $\lambda_C$ = 150 pct/s and processing rate of $\mu_C$. Each packet is reduced and forwarded to the Fog node, with maximum rate $\lambda_M$ = 150 pct/s, and average packet size of 10 bytes. The processing rate considered at Fog node is also 115200bps, thus $\mu_M$ = 1440 pct/s. These queues are then reduced into one queue with input $\lambda_{CM}$ = 150pct/s with packet size of 35 bytes, and $\mu_{CM}$ = 411 pct/s.

Lastly, we consider the Fog to Mist queue for feedback control packets, that are 5 bytes long with $\lambda_{FM}$. The Mist node microcontroller will process the control packets, and forward to antenna the ZigBee packets that are 25 bytes long as $\lambda_{FC}$, thus, total feedback $\lambda_{FMC}$ average packet size will be 30 bytes.

Accordingly, we redo the estimate savings calculations for $E\{T_{CM}\}$ with and without feedback. Results are presented in table \ref{T_CM_lambda_feedback}. In the case without feedback, maximum T allowed is T=2.82s, resulting in 58.5\% savings, or 0.5\% less than without the Mist layer.

In the smart condominium fire alarm system, with the Mist layer and feedback $0.01\mu_{CM}$, $\lambda_{FMC}$ will be 4 pcts/s. IoT devices have an estimated maximum energy savings of 58.4\%, considering its working cycle with maximum T=2.81 sleep time, and average waiting time E\{T\}=181ms for packets queue processing delay, when scaling to 300 devices. 

Results in table \ref{T_lambda_feedback} and \ref{T_CM_lambda_feedback}, show maximum savings with and without feedback and Mist layer, estimating about 5\% difference when adding a Mist layer. We must then consider additional ways to save energy. 

\begin{table}[!ht]
\centering
\caption{IoT Network with Mist and Feedback Savings S(\%) }

\resizebox{.49\textwidth}{!}{
\begin{tabular}{cccccccc}
Feedback & \textbf{T} & $\lambda_C$ & $\lambda_{CM}$ & $E\{T_C\}$(s) & $\rho_C$ & TT(s) & S(\%)\\ 
\hline
No feedback & 2.82 & 62.2 & - & 0.1785 & 0.1514 & 2.998 & 58.5   \\ 
$0.01\mu_{CM}$ & 2.81 & 62.4 & 4 & 0.1810 & 0.1532 & 2.991 & 58.4   \\ 
$0.05\mu_{CM}$ & 2.8 & 62.5 & 21 & 0.1908 & 0.1603 & 3.000 & 58.3   \\ 
$0.10\mu_{CM}$& 2.79 & 62.6 & 41 & 0.2038 & 0.1693 & 2.994 & 58.2   \\ 

\end{tabular}}
\label{T_CM_lambda_feedback}
\end{table}

\subsection{Autonomic dynamic duty cycle with Home Scheduling System}

Now, we consider that IoT devices are in standard operating mode at times of the day that the residents are in the apartments. The standard operating mode follows the 5-second limit, being 2 seconds active and 3 seconds sleep limit to inform a local emergency fire state, thus alerting inhabitants in time to act or flee. For periods when residents are not at home, we then allow T values to be higher than the initial 3 seconds limit, being defined as \textbf{Long Sleep (LS)} state, while also disregarding the system's queue delay time. 

Considering a long sleep of 4 seconds, in a total duty cycle of 6 seconds, and with 10 cycles per minute, the one second difference above sleep limit T=3, would make little difference for residents' safety if they are away from the premises. With 2 seconds active and 10 seconds sleeping, the LS cycle would be 12 seconds, and in one minute, there would be 5 cycles, thus allowing for more significant device energy savings. If the IoT device detects an emergency fire state while the resident is out, an automatic message is sent from the Fog server to the resident's phone.

\begin{figure}[!htb]
\centering
\includegraphics[width=0.3\textwidth]{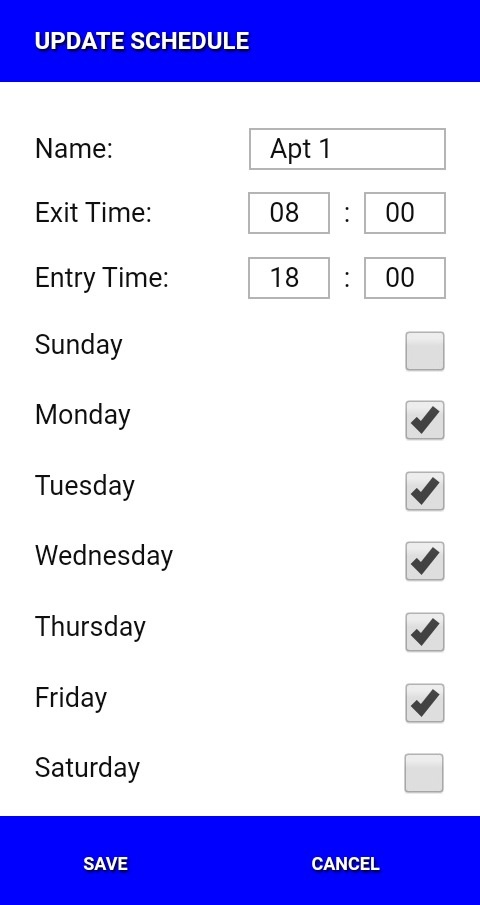}
\caption{Fog Web System's Home Schedules Update Screen}
\label{fig:schedule_update}
\end{figure}

For the 300 apartments in the condominium, each resident may have different daily schedules for leaving and returning home, such as going to work, school, gym or any recurring activities. To determine when the apartments are empty, we then propose and develop a scheduling system, where each user registers its own schedules with the apartment number, exit and entry times, as well as which days of week it should run, shown in Fig. \ref{fig:schedule_update}.

Therefore, we estimate daily IoT energy consumption considering users' time schedules to determine when residents are at Home (H) or Away (A). In each schedule, exit (ex) and entry (en) times are informed, and its daily difference are the times when residents are away, in Eq. \ref{eq:nothome}. Thus, the times residents are home, in Eq. \ref{eq:home}, is the difference of A from the rest of the day:

\begin{equation}
\label{eq:nothome}
   \textrm{A} = \sum{(\textrm{en} - \textrm{ex})} 
\end{equation}

\begin{equation}
\label{eq:home}
   \textrm{H} = 24\textrm{hrs} - \textrm{A} 
\end{equation}

To estimate the condominium IoT fire alarm devices energy savings difference, with Long Sleep (LS), we then simulate schedules for $n=300$ apartments, arranged into 5 groups. In each group the apartments have the same average times for exiting and returning home, defined in table \ref{tab:cenarios}.

Each of the apartments from group 1 have two different schedules. In the first schedule the resident will leave the apartment at 8:00 in the morning, and will return at 11:30am, being away from home for three and a half hours. In the second schedule the resident will leave at 12:30pm and return at 06:00pm (18:00), being away from home for five and a half hours, totaling $A=9:30$ hours away.

Then, A and H times of each group are calculated with Eqs. \ref{eq:nothome} and \ref{eq:home}. Time results are divided by 24hrs to calculate A and H daily percentages, shown in table \ref{tab:cenarios}.

\begin{table}[!htb]
\centering
\caption{Simulated 5 Groups Exit (ex) and Entry (en) Schedules, total Away (A) and Home (H) times in hours (hrs), and daily A\% and H\% proportion}
\label{tab:cenarios}
\resizebox{0.48\textwidth}{!}{
\begin{tabular}{cl|ccccc}
\multicolumn{2}{c|}{\textbf{Groups}}           & \textbf{1}   & \textbf{2}  & \textbf{3}  & \textbf{4}  & \textbf{5}  \\ \hline
\multirow{2}{*}{\textbf{1st schedule}} & ex  & 08:00        & 06:00       & 11:00       & 7:00        & 17:00       \\
                                       & en & 11:30        & 13:00       & 16:00       & 9:30        & 24:00       \\ \hline
\multirow{2}{*}{\textbf{2nd schedule}} & ex  & 12:30        &             & 20:00       & 13:00       &    00:00         \\
                                       & en & 18:00        &             & 22:00       & 15:40       &   05:00          \\ \hline
\multirow{2}{*}{\textbf{3rd schedule}} & ex  &              &             &             & 18:00       &             \\
                                       & en &              &             &             & 20:50       &     \\
                                       \hline
\multicolumn{2}{c|}{A (hrs)} & 9:30 & 7:30 & 7:00 & 8:00  & 12:00 \\
\multicolumn{2}{c|}{H (hrs)} & 14:30 & 16:30 & 17:00 & 16:00 & 12:00\\ \hline
\multicolumn{2}{c|}{A (\%)} & 39.58 & 31.25 & 29.17 & 33.33  & 50 \\
\multicolumn{2}{c|}{H (\%)} & 60.42 & 68.75 & 70.83 & 66.67 & 50  
\end{tabular}}
\end{table}

Hence, the estimated condominium daily energy savings E, with Mist layer, for \textbf{n} apartments, is the sum of times residents are at home, multiplied by maximum estimated $T$ savings in table \ref{T_CM_lambda_feedback}, and times the residents are away from home, multiplied by $LS_{Savings}$ in table \ref{tab:lssavings}, shown in Eq. \ref{eq:dailyconsumption}. 

\begin{equation}
\label{eq:dailyconsumption}
    \textrm{E} = \sum_{i=1}^{n}{((H_i * T_{Savings}) + (A_i * LS_{Savings}))}
\end{equation}

To demonstrate calculations numerically, residents from group 1 will be at home for 14:30 hours a day, or 60.42\% of the time, with maximum T=2.81 or 58.4\% savings. While away from home 9:30 hours a day, or 39.58\% of the time, and maximum sleep considered is LS=4 seconds, or 66.66\% savings, with results given in Eq. \ref{eq:dailyconsumptionexample}.

\begin{equation}
\label{eq:dailyconsumptionexample}
    E_{1} = (60.42\% * 58.4\%) + (39.58\% * 66.66\%) = 61.67\%
\end{equation}

Total daily savings E for group 1, with LS = 4 seconds, is 61.67\% when compared with not using sleep mode T=0. This results in 3.27\% extra savings (ES) when compared with the sole daily use of standard operating mode when T=2.81. We calculate E for all groups, with results shown in table \ref{tab:cenarios2}. We also calculate Total A (TA) and Total H (TH) times, in hours, for each group, by multiplying the number of apartments by A and H values in table \ref{tab:cenarios}. 

We then calculate proportional A(\%) for each group, with results indicating that group 1 is responsible for 53.65\% of TA, being the group with the highest weight. Group 5 is the group with least apartments, 20, but its A\% of 9.04\% is higher than group 3 with 30 apartments and A\% of 7.91\%. To calculate the total condominium energy savings, we multiply E of \textbf{n} groups \textbf{g} by its respective A\%, and sum its results, shown in Eq. \ref{eq:extrasavingsls4}. 

\begin{equation}
\label{eq:extrasavingsls4}
    \textrm{Condominum E} = \sum_{g=1}^{n}{(E_{g} \times A_{g}\%) }
\end{equation}

\begin{table}[!htb]
\centering
\caption{Estimated Condominium Extra Savings $ES_g$(\%) with 4 Seconds LS, with Total Time Away ($TA_g$), Total Time Home ($TH_g$), Away Proportion ($A_g(\%)$) and Energy Savings($E_g(\%)$) per Group g.}
\label{tab:cenarios2}
\resizebox{0.5\textwidth}{!}{
\begin{tabular}{c|c|c|c|c|c|c}
\textbf{Groups} & \textbf{Apts} & \textbf{$TA_g(hrs)$} & \textbf{$TH_g(hrs)$} & \textbf{$A_g(\%)$} & $E_g$(\%) & \textbf{$ES_g$(\%)} \\ \hline
\textbf{1} & 150 & 1425 & 2175 & 53.67 & 61.67 & 3.27 \\
\textbf{2} & 40  & 300  & 660  & 11.30 & 60.98 & 2.58   \\
\textbf{3} & 30  & 210  & 510  & 7.91  & 60.81 & 2.41     \\
\textbf{4} & 60  & 480  & 960  & 18.08 & 61.15 & 2.75  \\
\textbf{5} & 20  & 240  & 240  & 9.04  & 62.53 & 4.13      \\
\hline
\textbf{Condo}  & \textbf{300}        & \textbf{2655}          & \textbf{4545}          & \textbf{100}           & \textbf{61.51}       & \textbf{3.11}                  
\end{tabular}
}
\end{table}

The results in table \ref{tab:cenarios2} indicate an average Condominium $E_{g}\%$ savings of 61.51\%. To estimate condominium extra savings with LS, we calculate with Eq.\ref{eq:extrasavingsls4} substituting $E_g$ by $ES_g$. Group 5 presents the highest extra savings $ES_5$ of 4.13\%. The condominium average extra savings is 3.11\% higher when applying LS=4 for times residents aren't home. We repeated calculations for total condominium E and ES while varying LS values, with results in table \ref{tab:lssavings2}.

\begin{table}[!ht]
 \centering
 \caption{Condominium energy consumption (kWh) difference (diff) per day, month, year, and extra energy savings E(\%) with LS(s) variation.}
 \label{tab:lssavings2}
\resizebox{.5\textwidth}{!}{
    \begin{tabular}{c|c|c|cc|cc|cc}
 \textbf{T(s)} &  \textbf{E(\%)} & \textbf{ES(\%)} & day & diff  & month & diff & year &  diff\\ \hline
 0 & 0 &  -  & 7.05 & - & 211.35  & - & 2536.22 & - \\
2.81 & 58.4  &  -     & 2.93 & 4.11 & 87.92 & 123.43 & 1055.07 & 1481.15 \\
\hline
LS(s)& - & - & - & - & - & - & - & - \\
\hline
4    & 61.51 & 3.11  & 2.71 & 0.22 & 81.35 & 6.57   & 976.19  & 78.88   \\
8    & 66.52 & 8.12  & 2.36 & 0.57 & 70.76 & 17.16  & 849.13  & 205.94  \\
10   & 67.78 & 9.38  & 2.27 & 0.66 & 68.10 & 19.82  & 817.17  & 237.90  \\
13   & 69.03 & 10.63 & 2.18 & 0.75 & 65.46 & 22.47  & 785.47  & 269.60  \\
18   & 70.29 & 11.89 & 2.09 & 0.84 & 62.79 & 25.13  & 753.51  & 301.56  \\
28   & 71.54 & 13.14 & 2.01 & 0.93 & 60.15 & 27.77  & 721.81  & 333.26  \\
58   & 72.80 & 14.40 & 1.92 & 1.01 & 57.70 & 30.22  & 692.39  & 362.68  \\ \hline
\end{tabular}   
}
\end{table}

We also calculated the consumption differences when varying sleep time T=2.81, and long sleep LS times, in table \ref{tab:lssavings2}. The difference in energy consumption between T=0 and T=2.81 is 58.4\%, or up to 1481.5 kW per year. When comparing savings of T=2.81 with LS=4, results in 3.11\% extra savings, being 220W per day, 6.57 kW per month or 78.88 kW per year. With LS=8 the condominium is able to save up to 8.12\%, or 205.94 kW per year. However, increasing LS time does not result in proportional extra savings, for example, with LS=28 seconds its estimated ES is 13.14\%, and with LS=58 seconds the ES is 14.40\%. The 30 seconds difference between LS=28 and LS=58 only resulted in a 1.26\% efficiency gain.

\section{Related Work}
\label{sec:background}

We acknowledge some related work  around the utilization of Fog Computing in  Smart Environments and position our work in areas where further research is needed. 

Existing IoT-based energy management projects focus on radio data transmission in order to increase efficiency. Our argument is that energy efficiency is attained by orchestrating controlling radio, microcontroller, and sensors/actuators modules during active mode, as well as turning off unnecessary modules during sleep mode. This feature is overlooked in current solutions, as analyzed above. 

Ghidini and Das \cite{Ghidini2012b, Ghidini2012a} introduced $E^{2}$ home to calculate multiple homes' energy consumption, by reading the energy meter information from Smart Meter Texas (SMT) Web site. $E^{2}$ home  displayed high energy usage patterns, for times residents were at home or away, where residents’ locations were determined from Android smartphones. To achieve energy savings, the user had to analyze the energy consumption history, and take actions to adjust their behavior, to minimize wasteful consumption.

Bhilare and Mali \cite{Bhilare2016} developed an AMS IoT device connected to the home energy meter, and relay modules connected to a light, a fan, and an air-conditioner, being able to manage appliances and energy consumption individually. The user was able to set in the system, a home energy meter threshold, to inform residents if the consumption limit had been exceeded, thus, user was responsible to take action in order to minimize energy waste. 
 
Ku, Park and Choi \cite{Ku2017} realized an IoT energy management platform, using microgrid, for a community of 110 homes. In this solution, each home was controlled by a Hub device, which interconnected with the management office center server. Each hub, controlled the home microgrid, connected to electricity and gas meters, lights control, and Heating Ventilation and Air Conditioning (HVAC). Outlets are also connected to the microgrid, to individually manage household appliances. Management office was responsible to analyze each home's energy usage history and trends, and feedback information to help residents reduce energy wastage.

Chen, Azhari, and Leu \cite{Chen2018}, proposed a Fog and IoT electrical consumption monitoring system, with the use of ZigBee smart sockets, for a smart home. Home outlets were considered as IoT nodes which allowed energy monitoring of appliances connected to them. A system performance evaluation was carried out, for a period of 15 minutes, in order to compare the latency time when using a local Fog server, and a Cloud server. The observed latency, when using only the Cloud, varied between 700ms and 2s, while, when sending data to be processed locally by Fog, the latency time was below 10ms. This approach can effectively address the data amount issues and reduce the response latency to secure the safety for the family members in real-time processing needed situations.

The system introduced by Agyeman, Al-Waisi and Hoxha \cite{Agyeman2019} consisted of a microgrid managed by an IoT device with a non-invasive current sensor, and multiple relays, acting as smart outlets. The IoT device was responsible for managing the household appliances by switching them on and off. The current sensor read each appliances' electricity consumption, and forwarded data to a Fog device through an Ethernet connection. To help residents reduce energy consumption, authors also uses threshold notification if consumption budget is exceeded.

Hijawi, Gastli, Hamila, Ellabban and Unal \cite{Hijawi2020} proposed a large scale IoT energy management system for a school building, with the use of smart WiFi outlets. The system considered each outlet as an individual IoT smart meter unit, with current sensor and relay modules. IoT data was forwarded to be stored in a Cloud server, where individual energy consumption history was displayed. To achieve efficient energy usage, user was responsible to send power control commands to the IoT units, at different times of the day, based on the data monitored and retrieved from the server over time.

Beaudaux, Gallais \& No{\"{e}}l~\cite{Beaudaux2013} produced a Heterogeneous radio preamble-sampling MAC duty-cycling, for energy-efficient Internet of Things deployments. In such protocols, nodes cyclically activated its radio to sample the channel for incoming messages, being referred to as Low-Power-Listening or LPL. For most existing solutions, LPL is homogeneous for network nodes, thus, it is proposed a strategy was proposed where different LPL configurations cohabited in a single Network of Things, corresponding to each node's role in the network, for dense large scale networks. Performance was evaluated through simulation, where results showed an average energy reduction of 61\%, with a drawback of higher loss-rate and network delay.

Jayakumar, Raha and Raghunathan \cite{Jayakumar2014} promoted an ultra-Low power sleep mode, with SRAM data retention for embedded microcontrollers, named Hypnos. The paper focused on microcontroller's energy consumption while in idle mode. Hypnos considered that the on-chip SRAM is capable of 100\% data retention, even at much lower supply voltage than regular operating voltage. Thus, performing extreme voltage scaling, when the microcontroller was in idle mode, resulted in decreasing idle power consumption by 75\%, when compared to conventional implementation.

In Kolios, Ellinas, PanayIoTou, Polycarpou \cite{Kolios2016}, authors developed an Energy Efficient Event-Based Networking for IoT, considering the minimum number of triggering events necessary to achieve the desired tracking accuracy and/or to guarantee control stability, in order to achieve the maximum network-wide energy efficiency. Energy savings was achieved by allowing communication circuitry to wake up only when particular events took place. The solution was tested through simulations, varying the number of receiving nodes and triggering events. the results indicated that, when the transmit time duration was in the order of a few hundred milliseconds to seconds, a gain of more than 50\% was observed compared to the nominal solution, and with lower transmission time, higher savings could be achieved.

In Lu, Kim, Xhafa, Zhou and Tsai \cite{Lu2017}, authors presented a method to reach 10-years of battery life for Industrial Internet of Things (IIoT), using a platform called the I3Mote. Authors measured the radio wake-up time, being responsible for around 20\% of packet transmission. Consequently, it was considered to use a faster start-up crystal oscillator to accelerate active/sleep transition for radio, sensors, and power-up times. On the network side, it is proposed to use Time Slotted Channel Hopping (TSCH) protocol, where the network coordinator sent periodic beacons and scheduled timeslots transmission among nodes. Each node shared a schedule, allowing it to know in advance when to turn on or off its radio. Thus, by increasing duty cycle to 500 seconds, 11.5 years battery lifespan could be achieved.

Al-Janabi and Al-Raweshidy \cite{Al-Janabi2019} proposed a hybrid time division multiple access (TDMA)-carrier sense multiple access, with a collision avoidance mechanism (CSMA/CA) MAC protocol, called HSW-802.15.4, for large scale with high density IoT networks. The scheduling sleeping technique was adjusted according to high collision or channel failure conditions, determining the group of devices that were active per timeslot. The proposed protocol efficiently utilized the energy of the nodes and dynamically adapted the sleep/wake-up periods according to the variance in the network loads. Simulation results showed that HSW-802.15.4 can efficiently utilize the radio energy up to 60\%.

Sampaio, Jesus, Boing and Westphall \cite{sampaio2019} developed a HEMS and a real time fire alarm system, with Fog and battery powered IoT device. The IoT device had temperature, gas/smoke, and flame sensors, to monitor the environment conditions, and a ZigBee antenna to send data to the Fog server. The Fog EMS remotely managed the IoT device altering its sleep time, considering the fire alarm real time restriction, thus, altering its duty cycle and energy consumption. Battery lifespans were calculated with duty cycle variation, and emergency state. With a maximum sleep duty cycle of 3 seconds, it was possible to reduce up to 59.99\% of IoT energy consumption.  

Perkovi{\'{c}}, Damjanovi{\'{c}}, {\v{S}}oli{\'{c}}, Patrono and Rodrigues \cite{Perkovic2020} reviewed strategies for reducing the consumption of Low Power Wide Area Networks (LPWAN) battery-powered devices. The authors developed a battery powered IoT device, with an Arduino Pro Mini, low power environment sensors, and a LoRaWAN radio, for smart environments with large distance connection from IoT to the gateway node. To minimize the device's energy consumption, all LEDs and voltage regulator were removed from the main board. Another strategy for energy savings was the use of Low-Power library for Arduino, named deep sleep, that could reduce microcontroller sleep consumption below 1$\mu$A. The IoT device was developed with an RTC module, responsible for waking up the device from very long deep sleep cycles of 20 minutes.

Amirinasab, Shamshirband, Chronopoulos, Mosavi and Nabipour \cite{AmirinasabNasab2020} designed a lightweight clear channel assessment (LW-CCA), as an extension of ContikiMAC, a low power radio duty cycling protocol available in Contiki OS based on the low power listening (LPL) mechanism. LW-CCA reduces the percentage of Radio Duty-Cycles in false WakeUps and idle listening, by using dynamic received signal strength indicators (RSSI) status check time. Simulation results showed that LW-CCA reduced about 8\% energy consumption in nodes, while maintaining up to 99\% of the packet delivery rate (PDR).

We conclude that modes of the existing EMS solutions focused on a combination between data from energy meters, individual appliances, and historical data analysis. By configuring energy consumption thresholds, residents could receive automatic information messages as recommendations to decide on corrective actions. However, existing approaches still rely heavily on user affirmative action in order to obtain energy savings. Hence, there is an opportunity to contribute with solutions able to manage IoT device operations by controlling individual modules, thus defining IoT duty cycles. 

\section{Conclusions}
\label{sec:conclusion}

We concluded that there is a significant opportunity for energy saving by properly balancing the demand for quality of service and sleep time. Our tests indicate energy savings of 61.51\% when Long Sleep (LS) is set to the standard of 4 seconds. The increase of LS setting does not have a proportional relation to increase savings. If we set LS to 58 seconds, it leads to an extra energy savings of 14.4\%. 

The proposed Central IoT (CIoT) provides an innovative view for a Autonomic IoT energy management system based on Fog and Mist Computing. The solution performs IoT device power management, with consumption estimate calculations in kWh, correlating variations of environment (temperature, smoke and flames), network load, and residents behavioral information for home occupation (daily schedules), realizing autonomic adjustments to regulate energy utilization.

This proposal is relevant for large scale IoT energy consumption environments as well as for energy generation and distribution planning, as it is possible to estimate the fluctuation in demand based on the configured sleep times in large IoT deployments. 

In future work, we intend to add a condominium IoT home access control with RFID subsystem, in order to obtain more precise information about resident's exit and entry times, and other forms of parameter provisioning \cite{assuncao2017content}. We will also look into security issues and intend to investigate the use of Intrusion Detection \cite{vieira2019autonomic, schulter2006towards, schulter2006grid} and others to safeguard the IoT control environment. Finally, we seek to include methods of Computational Intelligence, Machine Learning, and Distributed AI \cite{assuncao2004grids, parker2007distributed} to estimate calculations, considering parameters RFID-based presence identification, exit and entry times, environment conditions, the average waiting queue time variation, and others.

\section{Acknowledgments}

This work was partially supported by the Research and Innovation Support Foundation of the State of Santa Catarina (FAPESC) under grant 23038.013359/2017-71, the Coordination for the Improvement of Higher Education Personnel(CAPES), and National Council for Scientific and Technological Development (CNPQ).

\bibliographystyle{elsarticle-num}
\bibliography{mybib}
\end{document}